\documentclass[aps,prb,twocolumn,superscriptaddress]{revtex4-1} 

\usepackage{color}

\usepackage{graphicx}
\usepackage{epstopdf}
\usepackage{color}
\usepackage{amssymb,amsfonts,amsmath}
\usepackage{multirow}

\begin{document}

\author{Friederike Schmid}
\email{schmidfr@uni-mainz.de}
 \affiliation{Institut f{\"u}r Physik, Johannes Gutenberg
Universit{\"a}t  Mainz, Staudingerweg 9, 55128 Mainz, Germany}

\author{Bing Li}
 \affiliation{Institut f{\"u}r Physik, Johannes Gutenberg
Universit{\"a}t  Mainz, Staudingerweg 9, 55128 Mainz, Germany}

\title{Dynamic self-consistent field approach for studying 
kinetic processes in multiblock copolymer melts}






\begin{abstract}
The self-consistent field theory is a popular and highly successful
theoretical framework for studying equilibrium (co)polymer systems at
the mesoscopic level. Dynamic density functionals allow one to use this
framework for studying dynamical processes in the diffusive,
non-inertial regime.  The central quantity in these approaches is the
mobility function, which describes the effect of chain connectivity on
the nonlocal response of monomers to thermodynamic driving fields.  In a
recent study [Mantha et al, Macromolecules 53, 3409 (2020)], we have
developed a method to systematically construct mobility functions from
reference fine-grained simulations. Here we focus on melts of linear
chains in the Rouse regime and show how the mobility functions can be
calculated semi-analytically for multiblock copolymers with arbitrary
sequences without resorting to simulations. In this context, an accurate
approximate expression for the single-chain dynamic structure factor is
derived. Several limiting regimes are discussed. Then we apply the
resulting density functional theory to study ordering processes in a
two-length scale block copolymer system after instantaneous quenches
into the ordered phase. Different dynamical regimes in the ordering
process are identified: At early times, the ordering on short scales
dominates; at late times, the ordering on larger scales takes over.  For
large quench depths, the system does not necessarily relax into the true
equilibrium state.  Our density functional approach could be used for
the computer-assisted design of quenching protocols in order to create
novel nonequilibrium materials.
\end{abstract}

%
%


 \newcommand{\ud}{{\rm d}}
 \newcommand{\ue}{{\rm e}}

 \newcommand{\Rg}{R_g}

 \newcommand{\qq}{{\mathbf{q}}}
 \newcommand{\rr}{{\mathbf{r}}}
 \newcommand{\RR}{{\mathbf{R}}}
 \newcommand{\hmu}{{\hat{\mu}}}
 \newcommand{\hL}{{\hat{\Lambda}}}

 \newcommand{\matL}{\underline{\underline{\Lambda}}}
 \newcommand{\hmatL}{\hat{\underline{\underline{\Lambda}}}}
 \newcommand{\matG}{\underline{\underline{g}}}
 \newcommand{\matGG}{\underline{\underline{G}}}
 \newcommand{\vecrho}{\underline{\rho}}
 \newcommand{\vecjj}{\underline{{\bf j}}}
 \newcommand{\vecmu}{\underline{\mu}}
 \newcommand{\hvecmu}{\hat{\underline{\mu}}}
 \newcommand{\vecphi}{\underline{\phi}}
 \newcommand{\vecchi}{\underline{\chi}}
 \newcommand{\vecomega}{\underline{\omega}}
 \newcommand{\vecI}{\underline{I}}

 \newcommand{\tg}{\tilde{g}}
 \newcommand{\tn}{\tilde{n}}
 \newcommand{\tm}{\tilde{m}}
 \newcommand{\tq}{\tilde{q}}
 \newcommand{\tT}{\tilde{t}}

 \newcommand{\hinf}{H_\infty}
 \newcommand{\Happr}{H_{\mbox{\small appr}}}

 \newcommand{\twoscale}{A$_{5m}$B$_m$A$_m$B$_m$A$_m$B$_m$}

 \newcommand{\dir}{.}

 \newcommand{\RE}[1]{\textcolor{black}{#1}}


\maketitle

\section{Introduction}
\label{sec:introduction}

 
Block copolymers, i.e., polymers made of different chemically
incompatible units, are known to spontaneously self-assemble into a rich
variety of nanostructured patterns
\cite{Hong1981,Bates1996,Schacher2012}. The morphologies and dimensions
of these morphologies can be varied by tuning the molecular weight and
architecture of the constituent polymers.  This makes them interesting
for many applications such as drug delivery \cite{Liechty2010}, energy
conversion \cite{Peng2017a, Peng2017b}, or soft lithography
\cite{Black2007,Bates2014}, as well as for fundamental research.

Theoretically, the self-consistent field (SCF) theory has proved to be
a particularly valuable tool for studying self-assembled structures
and morphological phase diagrams \cite{Schmid1998, Matsen2002,
Fredrickson, Schmid2011}. In parameter regimes where thermal
fluctuations can be neglected, SCF models can often predict
equilibrium self-assembled structures at a quantitative level.
However, real materials often do not reach the true, fully ordered
equilibrium state on experimental time scales. Defects form during the
ordering process, which do not annihilate unless special techniques
are applied \cite{Mansky1998, Angelescu2004,Tsarkova2006, Yager2010,
Li2014,LiMu2015,Vu2018, Abate2019}.  Furthermore, intermediate states
may appear, which may be interesting by themselves and can stabilized
by crosslinking or freezing. The properties of these transition states
not only depend on the characteristics of the constituent molecules,
but also on the way the material is processed.  For these reasons,
considerable effort is also spent on studying the dynamics of BCP
ordering processes \cite{Fredrickson1996}.

\RE{SCF theories are often derived by field theoretical methods, {\em
i.e.}, first rewriting the partition function as a functional integral
{\em via} insertion of Delta functionals, and then applying a
saddle-point approximation. Similar approaches have recently been
taken to derive a dynamic SCF theory \cite{Fredrickson2014,
Grzetic2014}, starting from the Martin-Siggia-Rose functional for
Langevin dynamics\cite{MSR1973}.  Solving the resulting dynamic SCF
equations typically involves simulating an ensemble of independent
chains moving in a co-evolving self-consistent field
\cite{Grzetic2020}, similar to the 'self-consistent Brownian dynamics'
\cite{Saphiannikova1998, Saphiannikova2000, Ganesan2003},
'single-chain in mean field' \cite{Daoulas2006},  or 'MD-SCF'
simulation methods \cite{Milano2009} that have been used with great
success to study polymer systems in and out of equilibrium.}

\RE{Another popular class of dynamic extensions of the SCF theory is
the class of} dynamic self-consistent field or dynamic density
functional theories (DDFT), which combine the free energy functional
of the SCF theory with a diffusive dynamical model for the polymer
relaxation \RE{and do not require explicit chain simulations}.  The
generic form of dynamical equation of an inhomogeneous (co)polymer
system has the form \cite{Fraaije1993,Fraaije1997,Muller2005,Qi2017}
\RE{
\begin{equation}
\partial_t \vecrho(\rr,t)
  = \nabla_{\rr} \int \ud \rr' \: \matL(\rr,\rr')
    \nabla_{\rr'} \vecmu(\rr',t),
\label{eq:DDFT}
\end{equation}
}
where $\vecrho (\rr,t) = \big( \rho_{\alpha} (\rr,t) \big)$ denotes
the local densities at position $\rr$ and time $t$ of monomers of type
$\alpha$ in vector notation, $\matL(\rr,\rr')= \big(\Lambda_{\alpha
\beta} (\rr,\rr') \big)$ a mobility matrix, and $ \vecmu(\rr') = \big(
\mu_{\beta} (\rr') \big)$ is derived from the SCF free energy
functional $F\{\vecrho\}$ via $\mu_{\beta}(\rr',t) = \delta F/ \delta
\rho_{\beta}(\rr',t)$.  Hydrodynamics can be included by adding a
convective term \cite{Maurits1998} to (\ref{eq:DDFT}) and combining it with a
dynamical equation for fluid flow \cite{Zhang2011,Heuser2017}.
 
The mobility matrix relates the local thermodynamic force
$(-\nabla_{\rr'} \vecmu(\rr',t))$ on monomers at position $\rr'$ to
the monomer density current at position $\rr$, taking into account the
effect of chain connectivity. It thus incorporates the information on
polymer dynamics, e.g., internal chain relaxation and possibly
entanglements. It should be noted that an ''exact'' mobility matrix
should also depend on frequency according to the Mori-Zwanzig theory
\cite{Zwanzig1961,Mori1965,Zwanzig}. A generalized dynamic RPA (random
phase approximation) theory that includes memory has recently been
proposed by Wang et al \cite{Wang2019}. The central assumption of Eq.\
(\ref{eq:DDFT}) is that one can describe inhomogeneous polymer systems
by an effective Markovian model which accounts for the multitude of
relaxation time scales in polymer systems in terms of a suitable
(effective) nonlocal mobility matrix.

The question is how to determine this mobility matrix. A number of
expressions have been proposed in the literature \cite{Kawasaki1987,
Kawasaki1988, Fraaije1993, Fraaije1997, Maurits1997,Qi2017}, which rely
on more or less heuristic assumptions. On the other hand, it was found
that not only the time scales, but also the pathways of self-assembly
may depend critically on the specific choice of the mobility matrix
\cite{Sevink2005,He2006}. In a previous paper, we have therefore
developed a more systematic approach, where the mobility matrix is
constructed in a bottom-up manner from the single chain dynamic
structure factor in particle-based reference simulations
\cite{Mantha2020}. We have tested it at the example of diblock
copolymer melts with lamellar ordering and shown that DDFT calculations based
on our approach can accurately reproduce the ordering and
disordering kinetics in these systems. In fact, the DDFT results and the
corresponding computer simulation data were found to be in similar
quantitative agreement than SCF predictions and computer
simulation data for equilibrium structures.

In Ref.\ \cite{Mantha2020}, the mobility matrix was determined from
fine-grained simulation data. However, if reliable theoretical
expressions for the single chain dynamic structure factor are available,
our approach can also be used to derive analytic or semi-analytic
expressions for the mobility matrix, without having to resort to
fine-grained reference simulations. The purpose of the present paper is
to provide such a description for melts of linear multiblock copolymers
in the Rouse regime, i.e., the regime where chains are not entangled. We
will first discuss the dynamic structure factor of Rouse copolymers and
present a highly accurate analytical approximate expression, which can
be used for efficiently calculating the mobility matrix of linear
multiblock copolymers with arbitrary block sequence. To illustrate our
approach, we will then apply the dynamic theory to a particularly
interesting multiblock copolymer melt with two competing length scales
\cite{Nap2004,Nap2006}, and show how the competition affects the
pathways of self-assembly and the resulting final structures.

\section{Theory}
\label{sec:theory}

We consider melts of Gaussian chains of total length $N$ in the Rouse
regime at total monomer density $\rho_0$. Single non-interacting chains
are characterized by their radius of gyration $\Rg$ and the chain
diffusion constant $D_c$, or, alternatively, the Rouse time $\tau_R =
\frac{2}{\pi^2}{\Rg^2}{D_c}$.  Monomers have different types $\alpha$,
and the monomer sequence along the chains is described by a function
$\vecchi(n/N)$, with $\chi_\alpha(n/N) = 1$ if monomer $n$ is of type
$\alpha$, and $\chi_\alpha(n/N)=0$ otherwise ($\sum_{\alpha}
\chi_\alpha(n/N) \equiv 1$). Knowing $\vecchi$, one can calculate the
overall fraction of monomers $\alpha$ in the chain $f_\alpha = \int_0^1
\ud \tn \: \chi_\alpha(\tn)$.  The free energy of the melt is described
by a free energy functional $F\{ \phi_\alpha(\rr,t)\}$, which depends on
the rescaled local densities of type $\alpha$ monomers,
$\phi_\alpha(\rr,t) = \rho_\alpha(\rr,t)/\rho_0$. In practice, we will
consider block copolymers made of two types of monomers A and B, with
Edwards-type interactions characterized by a Flory Huggins parameter
$\chi$ and a Helfand compressibility parameter $\kappa$
\cite{Helfand1971}, and use the SCF free energy functional describing
this class of systems.  The relevant equations are summarized in
Appendix \ref{app:scf}.

As in Ref.\ \cite{Mantha2020}, we will use reduced quantities 
$\vecphi = \vecrho/\rho_0$, $\hvecmu = N \vecmu$ 
and $\hmatL = \matL/\rho_0 N$ to simplify the notation. 
Eq.\ (\ref{eq:DDFT}) then takes the form
\begin{equation}
\partial_t \vecphi(\rr,t)
  = \nabla_{\rr} \int \ud \rr' \: \hmatL(\rr,\rr')
    \nabla_{\rr'} \hvecmu(\rr',t).
\label{eq:DDFT_scaled}
\end{equation}
The thermodynamic driving field $\hmu_\alpha(\rr,t) = \frac{N}{\rho_0}
\delta F/\delta \phi_\alpha(\rr,t)$ is derived from the SCF functional
of the copolymer system. The corresponding equations are given in
Appendix \ref{app:scf}. 

Following Ref.\ \cite{Mantha2020}, we approximate the mobility matrix by that
of a homogeneous reference system. This implies, in particular, that it
is translationally invariant, $\hmatL(\rr-\rr')$, hence we can
conveniently rewrite Eq.\ (\ref{eq:DDFT_scaled}) in Fourier space as

\begin{equation}
\partial_t \vecphi(\qq,t)
  = - \qq^2  \: \hmatL(\qq) \: \hvecmu(\qq,t)
\label{eq:DDFT_fourier}
\end{equation}
with the Fourier transform defined {\em via}
$ f({\qq}) = \int d \rr \: e^{i \qq \cdot \rr} \: f(\rr), \quad
 f(\rr) = \frac{1}{V} \sum_{\qq} e^{- i \qq \cdot \rr} f(\qq)$.

We determine $\hmatL(\qq)$ using the ''relaxation time approach''
developed in Ref.\ \cite{Mantha2020}, i.e., we calculate it from the
characteristic relaxation times of the single-chain dynamic structure
factor $\matG(\qq,t)$ in the reference system:  
 \begin{equation}
   \hmatL(\qq) = \frac{1}{k_B T N^2}\: 
     \matG(\qq,0) \: \matGG^{-1}(\qq)  \: \matG(\qq,0) 
   \label{eq:lambda}
  \end{equation}
 \begin{displaymath}
   \quad \mbox{with} \quad
  \matGG(\qq) = \frac{q^2}{N} \int_0^\infty  \ud t \: \matG(\qq,t).
 \end{displaymath}
This expression has been constructed such that the DDFT consistently 
reproduces $\matG(\qq,t)$ when used to study the relaxation dynamics
of a single tagged chain. Further details can be found in Ref.\
\cite{Mantha2020}. 

The central input quantity is thus the single chain dynamic structure, 
defined as
\begin{equation}
 \matG (\qq,t) = \frac{1}{N} \iint_0^N \!\!\! \ud n \: \ud m  \:
  \vecchi(n/N) \otimes \vecchi(m/N)
   \left\langle 
    \ue^{i \qq \cdot (\RR_n(t)-\RR_m(0))} \right \rangle,
 \label{eq:gqt}
\end{equation}
where $\langle \cdot \rangle$ denotes the configurational average over
all chain conformations, $\otimes$ the tensor product, and $\RR_n(t)$
gives the coordinates of monomer $n$ at time $t$. In Ref.\
\cite{Mantha2020}, we propose to measure $\matG(\qq,t)$ from reference
particle-based simulations.  Here, we take an alternative approach and
estimate it from the analytical solution for free Gaussian Rouse
chains. For homopolymers, an exact expression is available \cite{Doi},
which has been discussed extensively in the literature in various
limiting regimes \cite{deGennes1967,Doi,Wang2019}. The
generalization to block copolymers is straightforward (see Appendix
\ref{app:gqt_appr}).  However, using the resulting expression in the
above formalism is not easy, because it involves an infinite sum over
Rouse modes.  To overcome this problem, we have derived an
approximate expression, which avoids the sum, but still accurate
reproduces $\matG(\qq,t)$ over the whole relevant range of $\qq$ and
$t$.  The derivation can be found in the Appendix \ref{app:gqt_limiting}.  
The result is

\begin{widetext}
 \begin{eqnarray}
 \label{eq:g_appr}
 \matG(\qq,t) &=& \: N \:
   \iint_0^1 \!\!\! \ud \tn \: \ud \tm \:
   \vecchi(\tn) \otimes \vecchi(\tm) \:
   \tg_{\tn \tm}(q R_g, t/\tau_R)
   \\ \nonumber
   \mbox{with} \quad
 \tg_{\tn \tm}(\tq,\tT) &=& \left\{
 \begin{array}{ll}
 \exp\Big[{- \tq^2  |\tn - \tm|}\Big] \:
 \exp\Big[{-\tq^2 
 \sqrt{\tT} \Big(
  \Phi\big(\frac{\tn - \tm}{\sqrt{\tT}} \big)
  + \Phi\big(\frac{\tn + \tm}{\sqrt{\tT}} \big)
  + \Phi\big(\frac{2-|\tn - \tm|}{\sqrt{\tT}} \big)
  + \Phi\big(\frac{2-\tn - \tm}{\sqrt{\tT}} \big)
 \Big)\Big] 
 } 
  & : \tT < \tau^* \\
 \exp \Big[{-\tq^2 W(\tn)\Big] \: \exp\Big[-\tq^2 W(\tm)\Big] \:
 \exp \Big[- \tq^2 \frac{2}{\pi^2}
   \big( \tT - 2 \cos(\pi \tn) \cos(\pi \tm) \: \ue^{- \tT} \big)} \Big]
 & : \tT > \tau^*
 \end{array} \right.
 \end{eqnarray}
 \begin{equation}
 \label{eq:w_phi}
 \mbox{and} \quad
 W(\tn) = \tn^2 - \tn + \frac{1}{3},
 \qquad
 \Phi(y) := \frac{2}{\pi^2} \; \Big(
   \ue^{- \frac{1}{4} (\pi y)^2} \sqrt{\pi} 
   - \frac{\pi^2}{2} |y| \: (1 - \mbox{Erf} (\pi |y|/2))
  \Big).
 \end{equation}
\end{widetext}
where Erf is the error function, and the scaled crossover time is
set to $\tau^* = 1.7$. As demonstrated in Appendix \ref{app:gqt_appr},
the relative error of $\tg_{nm}$ with respect to the exact solution is
less than $2 (q \Rg)^2 \times 10^{-4}$ over the whole range of $\qq$
and $t$ (see Figure \ref{fig:gqt_appr} in the Appendix).

Eq.\ (\ref{eq:g_appr}) shows that the behavior of $\matG(\qq,t)$
features two time regimes: At small times $t < \tau^*$, the full
spectrum of Rouse modes contributes to the dynamic structure factor in
a collective manner that can be captured by a scaling function
$\Phi(y)$. At large times $t > \tau^*$, only the leading Rouse mode
contributes. In the limit $t \to \infty$, $\matG(\qq,t)$ assumes the
asymptotic behavior
\begin{equation}
 \label{eq:g_asym} 
 \matG (\qq,t) \stackrel{t \to \infty}{\longrightarrow} 
 N \: \ue^{- q^2 D_c t}  \: \vecI(q \Rg) \otimes \vecI(q \Rg)  
\end{equation}
\begin{displaymath}
 \quad \mbox{with} \quad
 \vecI(\tq) = \int_0^1 \!\!\! \ud \tn \:  \vecchi(\tn)
      \: \ue^{- \tq^2 \: W(\tn)}. 
 \end{displaymath}
 This equation can also be derived independently, see Appendix
 \ref{app:gqt_limiting}, Eq.\ (\ref{eq:gqt_limiting}). In the limit $t
 \to 0$ and $(q\Rg) \to \infty$, the double integral over $\tn$ and
 $\tm$ is dominated by the sharply peaked term $\ue^{- (q \Rg)^2 | \tn -
 \tm|}$, i.e., by contributions of monomers that are close along the
 chain, $n \approx m$.  In that limit, one obtains the scaling form
 \begin{equation}
  g_{\alpha \beta} (\qq,t)
    \stackrel {t \to 0 \atop (q \Rg) \to \infty}{\longrightarrow}
      \delta_{\alpha \beta} \: \frac{2 f_\alpha N}{(q \Rg)^2}  \:
        F\Big( (q \Rg)^2 \sqrt{t/\tau_R} \Big)
 \label{eq:g_scaling}
 \end{equation}
 with the scaling function $ F(x) = \int_0^\infty \ud u \: \ue^{- u -
 x\Phi(u /x) }$, where $\Phi$ is defined as in Eq.\ (\ref{eq:w_phi}). This
 corresponds to Eq.\ (4.III.12) in Ref.\ \cite{Doi},
 generalized to linear multiblock copolymers.

 Finally, at $t=0$, we have $\tg_{\tn \tm} = \ue^{- \tq^2|\tn -\tm|}$
 for all $\tq$. For linear multiblocks containing
 a set $\{ \alpha_i \}$ of blocks of type $\alpha$ with block length
 $N b_{\alpha_i}$, the integral (\ref{eq:g_appr}) then gives
\begin{widetext}
 \begin{eqnarray}
 \label{eq:gq0_aa}
 g_{\alpha \alpha}(\qq,0)& =& \frac{N}{(q \Rg)^4}  
   \Big\{ 2 \sum_{\alpha_i}
   \big( \ue^{- (q \Rg)^2 b_{\alpha_i}} \!\!
              - 1 + b_{\alpha_i} (q \Rg)^2 \big)
   + \!\!\!\! \sum_{\alpha_i, \alpha_j \atop i \ne j} \!\!
     \big( \ue^{- b_{\alpha_i} (q \Rg)^2}\!\! - 1 \big) 
     \big( \ue^{- b_{\alpha_j} (q \Rg)^2}\!\! - 1 \big) \:
     \ue^{- \Delta_{\alpha_i,\alpha_j} (q \Rg)^2 }  \Big\}
\\
 \label{eq:gq0_ab}
 g_{\alpha \beta}(\qq,0) &=& \frac{N}{(q \Rg)^4} \:
   \Big\{ \sum_{\alpha_i, \beta_j}
     \big( \ue^{- b_{\alpha_i} (q \Rg)^2} - 1 \big) \:
     \big( \ue^{- b_{\beta_j} (q \Rg)^2} - 1 \big) \:
     \ue^{- \Delta_{\alpha_i,\beta_j} (q \Rg)^2}  \Big\}
 \end{eqnarray}
\end{widetext}
 where $N\Delta_{\alpha_i \alpha_j}$ or $N\Delta_{\alpha_i \beta_j}$
 is the number of segments separating the blocks ($\alpha_i$,
 $\alpha_j$) or ($\alpha_i$,$\beta_j$), respectively.

Using these results, we can now apply Eq.\ (\ref{eq:lambda}) to
evaluate the mobility function. In the regime $t/\tau_R > \tau^*$,
the time integral can be evaluated analytically
\begin{eqnarray}
   \lefteqn{ \tq^2 \: \int_{\tau^*}^{\infty} \ud \tT \: \tg_{\tn
\tm}(\tq,\tT) } && \\ \nonumber
   & =&  \frac{\pi^2}{2} \: \tg_{\tn \tm}(\tq,\tau^*) \; \:
      f\Big( 
         2 \ue^{-\tau^*} \cos(\pi \tn) \cos(\pi \tm),
              \frac{2}{\pi^2} \tq^2 \Big) 
\end{eqnarray}
\begin{eqnarray*}
  \mbox{with} \quad
    f(u, \lambda) 
      &=& 1 -  u \lambda \int_0^1 \ud x \: x^\lambda \ue^{u \lambda(x-1)}
\\
      &=& 1 + \ue^{-u \lambda} (-u \lambda)^{-\lambda} 
        \: \gamma(1+\lambda,-u \lambda),
    \label{eq:ff}
\end{eqnarray*}
where $\gamma(1+\lambda,u)$ is the lower incomplete Gamma function.
The other integrals have to be computed numerically. 

It is possible to determine the limiting behavior of
$\hmatL(\qq)$ in certain cases: In the limit $q \to \infty$, the time
integral in Eq.\ (\ref{eq:lambda}) is dominated by small times $\tT$.
We can then use the scaling form (\ref{eq:g_scaling}) to evaluate
$\matGG(\qq)$, giving $G_{\alpha \beta}(\qq) = \frac{1}{D_c}
\delta_{\alpha \beta} \frac{4 f_\alpha N}{(q \Rg)^4} C$ with $C =
\frac{2}{\pi^2} \int_0^\infty  x F(x) \ud x = 3.587$.  The static
single chain structure factor in this limit is given by $g_{\alpha
\beta}(\qq) = \delta_{\alpha \beta} \frac{2 f_\alpha N}{(q \Rg)^2} $.
Putting everything together, we obtain
 \begin{equation}
 \label{eq:l_asym}
 \hL_{\alpha \beta}(\qq) 
 \stackrel{q \to \infty}{\longrightarrow}
 \delta_{\alpha \beta} \frac{D_c}{k_B T} f_\alpha.
 \cdot 0.279
 \end{equation}
 
 In the limit $q \to 0$, we specifically examine the total mobility
 $\hL_{\mbox{\small total}}(\qq) = \sum_{\alpha \beta} \hL_{\alpha
 \beta}(\qq)$, which corresponds to the mobility function for
 homopolymers. The relevant contribution to the time integral entering
 $\hL_{\mbox{\small total}}(\qq)$ stems from late times, thus we can
 replace $\matG(\qq,t)$ by the asymptotic expression
 (\ref{eq:g_asym}), resulting in $G_{\mbox{\small total}}(\qq) \to
 \frac{1}{D_c} (1-\frac{(q \Rg)^2}{3})$.  Furthermore, we have
 $g_{\mbox{\small total}}(\qq,0) \to N (1-\frac{(q \Rg)^2}{3})$ at $q
 \to 0$. Together, we obtain
 \begin{equation}
 \hL_{\mbox{\small total}} (\qq) 
 \stackrel{q \to 0}{\longrightarrow}
 \frac{D_c}{k_B T} 
 \Big( 1-\frac{(q \Rg)^2}{3} \Big),
 \label{eq:l_total_asym}
 \end{equation}
 which essentially reflects the diffusion of the whole chain.
 Unfortunately, a similarly simple expression for the asymptotic
 behavior of the individual components $\hL_{\alpha \beta}$ for block
 copolymers is not available, since they also include contributions
 from the internal modes, which relax on time scales of order
 $\tau_R$.

\begin{figure}[t]
\includegraphics[width=9cm]{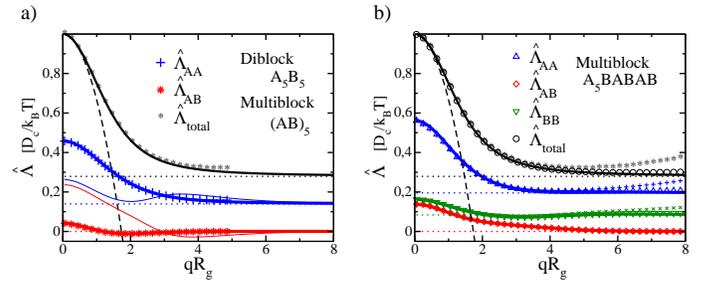}
\caption{Mobility functions $\hmatL_{\alpha \beta} (\qq)$ for (a)
symmetric AB diblock copolymers and (b) asymmetric multiblock
copolymers with sequence \twoscale.
Thick solid lines show the theoretical results from Eq.\ (\ref{eq:lambda}) 
with (\ref{eq:g_appr}). Symbols show simulation results from
particle-based simulations of chains with length 
$N=40$ (plus and stars) and $N=100$ (circles, triangles
and diamond). Dotted lines show the limiting behavior at
$q \to \infty$ according to Eq.\ (\ref{eq:l_asym}),
dashed line shows limiting behavior (\ref{eq:l_total_asym})
of the total mobility function at $q \to 0$.
For comparison thin lines in (a) also show mobility functions
for a symmetric multiblock with sequence (A$_m$B$_m$)$_5$).
}
\label{fig:mobilities}
\end{figure}

Figure \ref{fig:mobilities} shows examples of mobility functions for
three different types of linear multiblock copolymers containing two
monomer species A and B.  Also shown with dotted lines is the expected
asymptotic behavior at $q \to \infty$ (Eq.\ (\ref{eq:l_asym}), and with
black dashed lines, the expected asymptotic behavior of
$\hL_{\mbox{\small total}}$, at $q \to 0$ (Eq.\ (\ref{eq:l_total_asym}).

Figure \ref{fig:mobilities}a) focusses on sequences that are symmetric
with respect to exchanging A and B. The thick lines show the
mobility function for symmetric diblock copolymers, the thin
line the corresponding results for multiblock copolymers with sequence
(A$_m$B$_m$)$_5$. In the case of diblocks, we have also carried out
particle-based simulations of discrete Gaussian chains of length $N=40$
(symbols) for comparison. The data are in good agreement with the
theory. The behavior of the dynamic mobilities of diblock and multiblock
copolymers is qualitatively quite different: In diblock copolymers, the
blocks move largely independent from each other: The mobility component
$\hL_{AB}$ is close to zero in the whole range of $q$. In contrast, in
multiblock copolymers, the motion of blocks is highly cooperative at
small $q$ and they start to decouple only at $q\Rg > 1$. At $q
\to \infty$, they move independent from each other as expected, i.e.,
$\hL_{AB} \approx 0$. 

Figure \ref{fig:mobilities}b) shows the mobility function for a more
complicated asymmetric multiblock copolymer with sequence \twoscale.
It has a basic diblock structure, but one of the two blocks carries
itself a periodic multiblock sequence. The mobility function combines
features of the symmetric diblock and multiblock copolymer mobilities
shown in Figure \ref{fig:mobilities}a). The behavior of the A
component resembles that in regular diblock copolymers. The B
component tends to move cooperatively with the A component at small
$q$.  The joint mobility $\hL_{AB}$ is nonzero over a range of $q$
which is even wider than in the case of pure periodic multiblock
copolymers. The theoretical curves are again compared with simulation
data for chains of length $N=40$ (plus and stars) and $N=100$
(circles, triangles, diamonds). The agreement is excellent at
small $q$. For large $q$ the simulation data for shorter chains start
to deviate from the theory. This effect decreases with increasing
chain length. 

\section{Application to two-length scale block copolymers}
\label{sec:multiblock}

\begin{figure*}[t]
\centering
\includegraphics[width=12 cm]{\dir/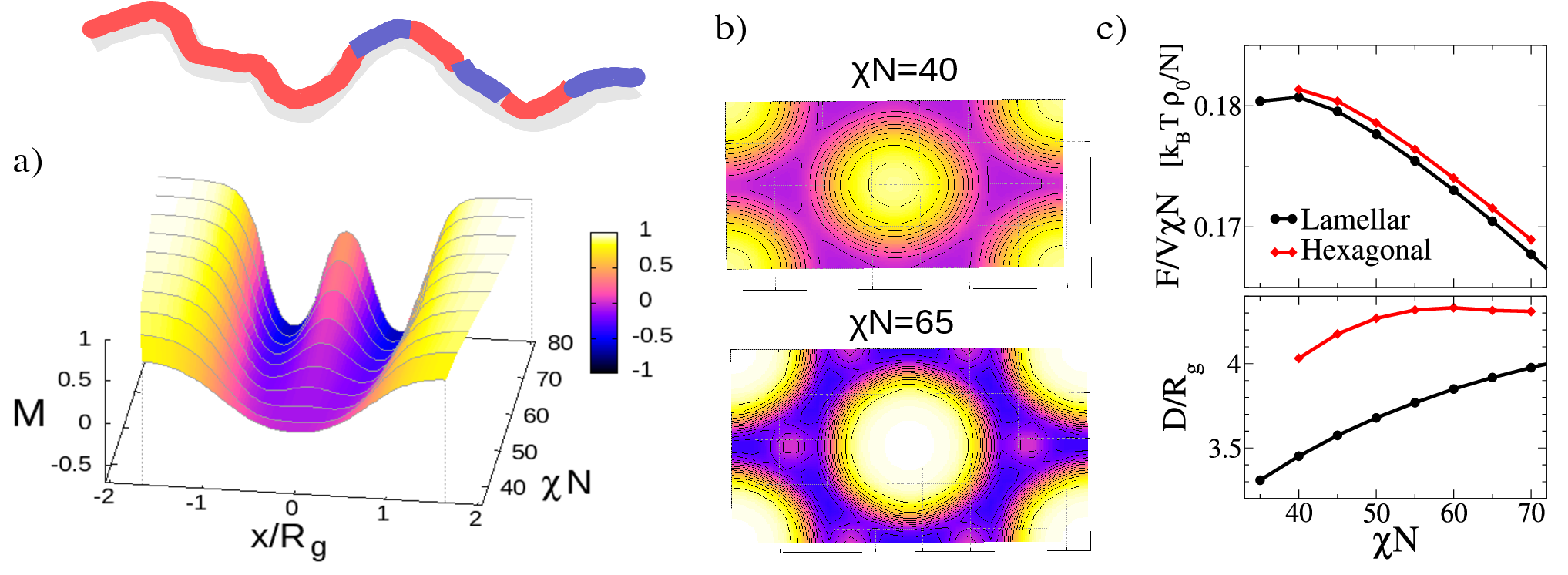}
\caption{Stable and metastable ordered phases in melts of multiblock
copolymers with sequence \twoscale.
(a) Order parameter profiles for stable lamellar structures as
a function of $\chi N$. A double lamellar structure emerges
for $\chi N > 42.5$.
(b) Order parameter maps for selected metastable hexagonal structures
for $\chi N=40$ and $\chi N = 65$. The color map is the same
as in panel (a).
(c) Free energy per volume (top) and periodic distance (bottom)
in lamellar and hexagonal phases as a function of $\chi N$.
In hexagonal phases, $D$ denotes distance between cylinders
}
\label{fig:scf}
\end{figure*}   

To illustrate our DDFT approach, we will now use it to study the
ordering kinetics of \twoscale block copolymer melts.  They belong to
a class of polymers with a two-length-scale molecular architecture,
which have attracted interest as promising candidates for responsive
materials \cite{Nap2004}. In particular, linear copolymers consisting
of one long uniform block and one periodic multiblock have been
studied in some detail, mostly by ten Brinke and coworkers
\cite{Nap2004,Nagata2005, Nap2006,Kriksin2008, Faber2012}.  At
sufficiently high $\chi N$, they form hierarchical patterns with small
structures embedded in larger ones. These two length scales should be
associated with two time scales, resulting in a complex ordering
kinetics.

To set the stage, we have determined the equilibrium structures of
\twoscale copolymer melts using SCF theory for $\chi N \in [35:80]$. The
results of the SCF analysis are summarized in Figure \ref{fig:scf}.  The
equilibrium structure is basically lamellar, but with a
lamellar-in-lamellar structure emerging at $\chi N \ge 42.5$,
characterized by an internal substructure inside the B domains. The
corresponding order parameter profiles are shown in Figure
\ref{fig:scf}a).  Here, the order parameter is defined as
$M=(\phi_A-\phi_B)/(\phi_A+\phi_B)$.  The lamellar phase competes with a
hexagonal phase, which also features substructures at higher $\chi N$.
Examples of order parameter maps of hexagonal structures corresponding
to local free energy minima are shown in Figure \ref{fig:scf}b). The SCF
free energy of the lamellar phase is always slightly smaller than that
of the hexagonal states, see Figure \ref{fig:scf}c). Along with the
minimum free energy per volume, Figure \ref{fig:scf}c) (lower panel)
also shows the periodic distance / lattice parameter of the minimum
structure.

Next we study the dynamic ordering process in this system using DDFT
calculations with the mobility function calculated in the previous
section (Figure \ref{fig:mobilities}b). The calculations were carried
out in two dimensions in periodic boxes of side length \RE{$17.2 \Rg
\times 15 \Rg$, using $86 \times 75$ grid points. These dimensions
were chosen such that, for every value of $\chi N$ studied here, at
least one side length was roughly commensurate with the equilibrium
lamellar distance and the lattice constant of the competing hexagonal
pattern.  The contour of the polymers was discretized with 100
''segments''. The time step was chosen $\Delta t = (0.2-1.0) \times
10^{-4} t_0$ depending on the system}, where the time unit is $t_0 =
\Rg^2/D_c$.  \RE{We found that the results do not depend on the
precise value of the time step, as long as the simulations were
stable. If the time step was too large, the numerical procedure to
determine the thermodynamic forces (see Appendix) failed, and we then
reduced the time step. In most systems, $\Delta t = 0.5 \times 10^{-4}
t_0$ was sufficient, but we had to set $\Delta t = 0.2 \times 10^{-4}
t_0$ in the most strongly interacting systems with $\chi N = 70$.} For
numerical reasons, we impose a frequency cutoff $\omega_c$, i.e., $(q
\Rg)^2 \hL_{\alpha \beta}$ may not exceed a cutoff value $\omega_c$.
This is necessary because $(q \Rg)^2 \hL_{\alpha \beta}$ in Eq.\
(\ref{eq:DDFT_fourier}) diverges at large $q$ for $\alpha=\beta$. The
frequency cutoff slows down local ordering processes on very short
time scales. Here, we use $\omega_c = 5/t_0$. We also did shorter
test runs on smaller systems with $\omega_c = 1.5/t_0$ (the value
found to be sufficient in our earlier work on diblock copolymers
\cite{Mantha2020}) and found that the rsults do not change
qualitatively.  The initial configuration is a homogeneous melt, to which
a small noise is added in order to initiate the ordering process.

\begin{figure}[t]
\centering
\includegraphics[width=9 cm]{\dir/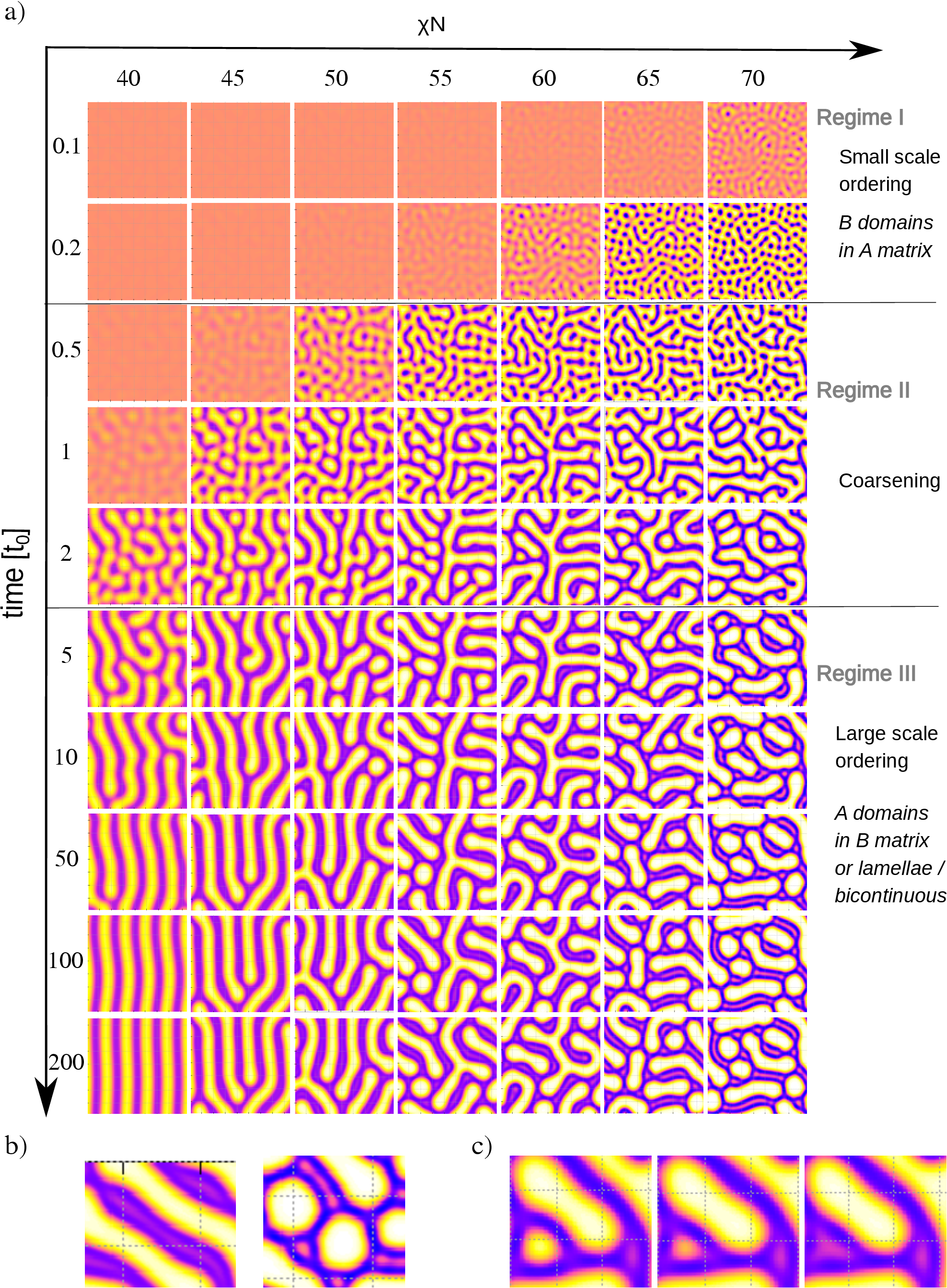}
\caption{a) Examples of ordering kinetics in \twoscale
diblock copolymer melts for different $\chi N$ and
identical disordered initial configuration. 
The color map is the same as in Figure 2.
b) \RE{Details} of final structures \RE{($t=200 t_0$) at $\chi N = 55$ (left)
and $\chi N = 70$ (right)} showing substructures in the B-domains.
c) \RE{Detail of a configuration} where a substructure emerged by 
dissolution of an A domain. \RE{Parameters are $\chi N = 60$ and
$t/t_0 = 2.0, 2.4, 2.5$.}  }
\label{fig:ddft_snapshots}
\end{figure}

Figure \ref{fig:ddft_snapshots}a) shows snapshots of melts during the
ordering process for different $\chi N$, starting from the same
initial configuration.  The ordering kinetics clearly reflects the two
scale character of the block copolymer. At early times $t < t_0$
(regime I), the local ordering is governed by the small characteristic
length scale. Around $t \sim t_0$, the structures start to coarsen
(regime II), until the second characteristic length scale is reached
around $t\sim (2-5) t_0$ (regime III). The actual kinetic ordering
pathway is governed by the interplay of these time-dependent ordering
scales with yet another time scale, the time required for A-B segregation,
which decreases with increasing incompatibility parameter $\chi N$. As
a result, the ordering process depends on $\chi N$. 

At low $\chi N$ (e.g., $\chi N =40$), the full segregation takes place
in regime III. A defective lamellar phase forms, which subsequently
orders by merging of A and/or B domains, resulting in an ordered
lamellar phase.  At intermediate $\chi N$, the segregation starts in
regime II and continues in regime III. Thus the system initially
orders on small scales, then coarsens by merging of B domains, but
merging of A domains is also possible.  The final structure is again
lamellar.  

Finally, at high $\chi N$ ($\chi N > 60$), the system segregates
already in the time regime I. It initially orders into small circular
$B$ domains, which then merge to form elongated connected structures.
Then coarsening sets in, which is first mediated by rearrangement and
further merging of B domains, and later by a thickening of B domains
associated with substructure formation inside them. In most cases,
A-rich substructures emerge spontaneously inside B domains.
Sometimes, we also observed that a larger A island dissolves into a
substructure (see Figure \ref{fig:ddft_snapshots}c)). Once regime III
is reached, the topology of the structures inverts from B domains in
an A matrix (in regime I) to A domains in a B matrix.
At late times, the A domains straighten out, but the basic topology of
the structure no longer changes. The final structure is characterized
by A domains with defined thickness but variable length, ranging from
circular to elongated.  Thus these final structures combine elements
of the equilibrium lamellar phase and the metastable hexagonal phase.
They are kinetically arrested and \RE{the topologies} do no longer
change. To model their further relaxation, one would have to add
stochastic thermal noise to the DDFT equations, following the lines
outlined in Ref.\ \cite{Mantha2020} (\RE{see next section}).  We
should however note that thermal fluctuation amplitudes in copolymer
melts are small \cite{Muller2005}, therefore we expect similar
long-lived structures to appear in real systems as well.

\section{Conclusions and Outlook}
\label{sec:conclusions}

To summarize, in this work, we have proposed a DDFT model for studying
kinetic processes of linear (multi)block copolymer melts in the Rouse
regime. The model builds on earlier work \cite{Mantha2020}, where we have
showed how to construct such models systematically from fine-grained
particle-based simulations. Here we use the same basic approach, but
calculate the central input quantity -- the mobility matrix --
semi-analytically from the theory of Rouse dynamics. One key ingredient
is an accurate approximate expression for the single chain dynamic
structure factor of Rouse chains, which can be applied at all times and
over a large wave vector range. 

\RE{DDFT models make more approximations and are less versatile than
self-consistent Brownian dynamics methods \cite{Fredrickson2014,
Grzetic2014}, which can also be used to study polymer systems far from
equilibrium where the use of SCF free energy functional is no longer
justified \cite{Saphiannikova1998}. On the other hand, they have the
advantage that they establish a natural connection to other dynamic
continuum theories such as Cahn-Hilliard theories. Furthermore, they
require relatively modest computational effort.  For example, the
calculations presented here (Figure 3) were run using own serial
code on an Intel Core i7-6700 CPU processor. The CPU time per time
step varied between 0.1 and 0.4 seconds, depending on the number of
iterations required to determine the thermodynamic driving force
self-consistently.  In total, the cost for simulating one Rouse time
$\tau_R = 2/\pi^2 t_0$ in our system of size 260 $\Rg^2$ was roughly
12 CPU minutes for $\chi N \le 65$, and 30 CPU minutes for $\chi N =
70$.}

To illustrate our DDFT approach, we have studied the ordering kinetics in a
melt of two-length-scale copolymers. The kinetic competition on
different length and time scales leads to an intricate interplay of
ordering processes and results in final structures that not
necessarily correspond to a true free energy minimum. Specifically, we
have studied situations where an initially disordered state was
instantaneously quenched into an ordered region. In that case, the
final structures strongly depend on initial small fluctuations and are
hard to control.  A better controlled ordering process might be
achieved by using a slower, well-defined and tunable quenching
protocol.  Experimentally, it is found that well-ordered two-scale
lamellar structures can be created by quenching the samples very
slowly \cite{Faber2012}. This is consistent with our calculations
where ordered structures are found to form when quenching into regions
with lower $\chi N$. \RE{We are not aware of published work on
non-equilibrium morphologies that can be obtained if samples are
quenched more rapidly. It would be interesting to compare them with
our numerical calculations.} We expect that it may be possible to
stabilize \RE{novel} structures when quenching with \RE{specially
designed} quenching \RE{protocols}, possibly combined with periodic
re-heating.  This could also be studied DDFT simulations and will be
an interesting direction for future work. 

\RE{In the present work, we have employed a deterministic DDFT model
that ignores thermal noise. Small thermal fluctuations can be included
in a straightforward manner by adding a stochastic current to the
DDFT equation, i.e., replacing  (\ref{eq:DDFT}) with
 \begin{equation}
 \partial_t \vecrho(\rr, t) = 
   \nabla_{\rr} \Big( \int \ud \rr' \matL(\rr,\rr') 
     \nabla_{\rr'} \vecmu(\rr',t) + \vecjj(\rr) \Big),
 \label{eq:DDFT_noise}
 \end{equation}
 where the components of $\vecjj(\rr,t)$ are Gaussian random variables
 with zero mean ($\langle j_{I \alpha} (\rr, t)\rangle = 0$) and
 correlations according to the fluctuation-dissipation theorem:
 \begin{equation}
 \langle j_{{}_I\alpha}(\rr,t) j_{{}_J\beta}(\rr',t') \rangle
  =  2 k_B T
    \delta(t-t') \: \Lambda_{\alpha \beta}(\rr,\rr') \: \delta_{{}_{IJ}}
\end{equation}
 (here $\alpha, \beta$ are monomer types and $I,J$ are cartesian
 coordinates).  Eq.\ (\ref{eq:DDFT_noise}) implicitly assumes that the
 SCF free energy functional (from which $\vecmu$ is derived) can be
 interpreted in the sense of a free energy landscape, which may 
 be questionable if fluctuations are large.  The relative
 amplitude of thermal noise is given by the inverse Ginzburg parameter
 \cite{Fredrickson, Qi2017} $C^{-1} = k_B T \: N/\rho_0 \Rg^3 \propto
 1/\rho_0\sqrt{N}$. In dense systems of polymers with high molecular
 weight, $C^{-1}$ is small. Thus fluctuations are small and can 
 be neglected in many cases, except when studying very soft modes
 (e.g., interfacial fluctuations) and/or dynamical pathways 
 that involve the crossing of free energy barriers.}

Our illustrative DDFT calculations were carried out in two-dimensions,
\RE{i.e., we have imposed uniformity in the third dimension. This was}
motivated by the fact that the relevant competing SCF structures of
our system are one- or two-dimensional.  However, in reality, the
initial structures \RE{will fluctuate in all three dimensions, e.g.,
one will find} be three-dimensional, e.g., small-scale spheres instead
of small-scale cylinders.  This will have to be elucidated by full
three dimensional calculations.

So far, the theory is restricted to linear multiblock copolymers, and we
have assumed that monomers are structurally similar, i.e., they have the
same flexibility and the same monomer friction. One goal of future work
will be to develop similar semi-analytic approaches for other polymer
architectures, for kinetically asymmetric copolymers, or (approximately)
for polymers beyond the Rouse regime. 

\vspace{6pt} 




\subsection*{Acknowledgments}
This research was funded by the German Science Foundation (DFG) via SFB TRR 146 
(Grant 233630050, project C1).



%



\begin{appendix}

\section{SCF equations}
\label{app:scf}

In the following, we briefly summarize the main SCF equations for our 
system. More detailed discussions of the SCF theory can be found,
e.g., in Refs.  \cite{Schmid1998,Matsen2002,Fredrickson, Schmid2011}.
In the SCF approximation, the free energy functional
$F\left[\{\phi_\alpha\left(\rr\right)\}\right]$ is expressed as
\begin{widetext}
 \begin{eqnarray}
 \label{eq:FE}
 \frac{F}{k_B T}
 = \frac{\rho_0}{N}
   \bigg\{ \int \ud \rr \Big[
      \chi N \: \phi_A\left(\rr\right)\phi_B\left(\rr\right)
    + \kappa N\left(\phi_A\left(\rr\right) + \phi_B\left(\rr\right)-1\right)^2
       \Big]
  - \int \ud\rr \: \vecphi\left(\rr\right) \cdot \vecomega\left(\rr\right)
     -V\ln Q\bigg\},
 \end{eqnarray}
\end{widetext}
 where $\vecphi=\{\phi_\alpha\}$ denotes the vector of normalized
 density field of monomers of type $\alpha$, $\vecomega =
 \{\omega_\alpha\}$ the corresponding vector of conjugate fields, and
 $Q$ the single chain partition function in the external fields
 $\vecomega$.  The conjugate fields are defined implicitly by the
 requirement
\begin{equation}
 \vecphi\left(\rr\right)
    = \frac{V}{Q}\int_0^{1} \ud \tn  \: \vecchi(\tn)
        \: q_f(\rr,\tn) \: q_b (\rr,1-\tn ),
 \label{eq:SCF1}
 \end{equation}
 where the chain propagators
 $q_f\left(\rr,s\right)$ and $q_b\left(\rr,s\right)$ are obtained
 from solving the differential equations
 \begin{eqnarray}
 \label{eq:prop1}
 \frac{\partial q_f(\rr,\tn)}{\partial \tn} 
   &=& \Rg^2 \nabla^2 q_f(\rr,\tn) 
     - \vecomega(\rr) \cdot \vecchi(\tn) \: q_f(\rr,\tn) \\
 \frac{\partial q_b(\rr,\tn)}{\partial \tn} 
   &=& \Rg^2 \nabla^2 q_b(\rr,\tn) 
     - \vecomega(\rr) \cdot \vecchi(1-\tn) \: q_b(\rr,\tn) 
 \label{eq:prop2}
 \end{eqnarray}
 with initial condition $q_{f,b}(\rr,0)=1$, and the single chain
 partition function $Q$ is given by
 \begin{equation}
   Q = \frac{1}{V}\int \ud \rr \: q_f(\rr,1)
     = \frac{1}{V}\int \ud \rr \: q_b(\rr,1).
 \label{eq:PartFn}
 \end{equation}
 The SCF equilibrium state is obtained by minimizing
 $F\left[\{\phi_\alpha\left(\rr\right)\}\right]$, leading to
 a second set of equations for $\vecomega(\rr)$:
 \begin{eqnarray}
  \omega_A^{\mbox{\tiny SCF}}\left(\rr\right)
      &=& \chi N\phi_B + 2\kappa N\left(\phi_A + \phi_B -1\right),
\\ nonumber
   \omega_B^{\mbox{\tiny SCF}}\left(\rr\right)
      & =& \chi N\phi_A + 2\kappa N\left(\phi_A + \phi_B -1\right)
 \label{eq:SCF2}
 \end{eqnarray}
 In DDFT calculations, these conditions are replaced by equations
 for the thermodynamic fields that drive the system towards
 equilibrium,
 \begin{equation}
 \hvecmu_\alpha(\rr)
   = \big(\vecomega^{\mbox{\tiny SCF}}(\rr)-\vecomega(\rr) \big)
 \end{equation}

 In SCF calculations, one must solve self-consistently the set of
 equations (\ref{eq:SCF1}) and (\ref{eq:SCF2}). In DDFT calculations,
 one must solve Eq.\ (\ref{eq:SCF1}) for the conjugated fields for given
 density fields $\phi_\alpha(\rr,t)$ in every time step. Both tasks
 require iterative methods. In the present work, we use Anderson 
 mixing \cite{Anderson1965} and a pseudospectral method for solving 
 the propagator equations (\ref{eq:prop1}) and (\ref{eq:prop2})
 \cite{Muller2005}.
 In the DDFT calculations, we require an accuracy of 
  $\max_{\alpha \rr}|\Delta \phi_\alpha(\rr)| < 0.001$ in every
  time step, where $\Delta \vecphi$ is the difference between
  the target density profile and the profile obtained from Eq.\
  (\ref{eq:SCF1}) with the given fields $\vecomega$. In the SCF
  calculations, we require
  $\max_{\alpha \rr}|\Delta \omega_\alpha(\rr)| < 10^{-6}$
  for the difference $\Delta \vecomega$ between the initial
  value of $\vecomega$ in an iteration step and the final value 
  computed from Eqs.\ (\ref{eq:SCF1}) and (\ref{eq:SCF2}).

\section{Single chain dynamic structure factor}
\unskip
\subsection{Late time behavior}
\label{app:gqt_limiting}

We consider the dynamic structure factor of single non-interacting
Gaussian chains in the Rouse regime.  Since all variables are Gaussian
distributed, Eq.\ (\ref{eq:gqt}) can be rewritten in the form
\begin{widetext}
 \begin{equation}
 g_{\alpha \beta} (\qq,t) 
   = \frac{1}{N} \iint_0^1  \!\!\! \ud n \: \ud m \:
   \chi_\alpha({\textstyle\frac{n}{N}})
   \: \chi_\beta({\textstyle \frac{m}{N}}) \:
     \exp\big(-\frac{1}{6} q^2
     \langle (\RR_n(t)-\RR_m(0))^2 \rangle\big) ,
 \label{eq:gqt_limiting}
 \end{equation}
\end{widetext}
 The trajectories $\RR_n(t)$ can be split up into the center of mass motion
 $\RR_c(t)$ and the relative motion $\rr_n(t)$,
 \begin{equation}
 \RR_n(t) = \RR_c(t) + \rr_n(t) \quad
 \mbox{with} \quad
 \RR_c(t) = \frac{1}{N} \int_0^N \!\!\! \ud n \: \RR_n(t).
 \end{equation}
 In the limit $t \to \infty$, the internal coordinates $\rr_n(t)$ and
 $\rr_m(0)$ are uncorrelated with each other and with $\RR_c$, i.e.,
 \begin{displaymath}
 \langle (\RR_n(t)-\RR_m(0))^2 \rangle   \approx
   \langle (\RR_c(t) - \RR_c(0))^2 \rangle
     + \langle \rr_n^2 \rangle + \langle \rr_m^2 \rangle.
 \end{displaymath}
 The first term describes the center of mass diffusion of the chain
 and obeys $\langle (\RR_c(t) - \RR_c(0))^2 \rangle = 6 D_c t$. 
 The second term describes the monomer fluctuations about the center 
 of mass. It can be calculated using the relation 
 $\langle (\RR_n - \RR_m)^2 \rangle = 6 |n-m| \Rg^2/N$, which
 is valid for Gaussian chains.  After some algebra, one obtains
 $\langle \rr_n^2 \rangle = 6 \Rg^2 \: \big( \textstyle (\frac{n}{N})^2  
 - \frac{n}{N}  + \frac{1}{3} \big) $.
 Thus the dynamic structure factor takes the asymptotic form
 $\matG(\qq,t) \approx 
 N \: \ue^{- q^2 D_c t}  \: \vecI(q \Rg) \otimes \vecI(q \Rg)$
 with
 \begin{equation}
 \quad \vecI(\tq) =
   \big\langle \int_0^N \!\!\! \ud n \:
       \vecchi({\textstyle\frac{n}{N}}) \: \ue^{i \qq \cdot \rr_n}
       \big\rangle
  = \int_0^1 \!\!\! \ud \tn \:
      \vecchi(\tn)\: \ue^{- \tq^2 \: (\tn^2 - \tn + 1/3)},
 \end{equation}
 which corresponds to Eq.\ (\ref{eq:g_asym}).

\subsection{Derivation of an approximation for $\matG(\qq,t)$}
\label{app:gqt_appr}

At finite $t$, the dynamic correlations due to internal modes can no
longer be ignored. The exact solution for $g(\qq,t)$ then involves an
infinite sum over all Rouse modes $p$. For homopolymers, the resulting
expression for homopolymers is derived, e.g., in Ref.\
\cite{Doi} (Appendix 4. III).  It can easily be generalized
to linear multiblock copolymers, giving
 \begin{eqnarray}
 \label{eq:g}
 \matG(\qq,t)
 &=& \ue^{- q^2 D_c t}
   \: N \:
 \iint_0^1 \!\!\! \ud \tn \: \ud \tm \: 
 \vecchi(\tn) \otimes \vecchi(\tm) \: 
 \\ \nonumber && \times
   \exp \Big[- (q \Rg)^2 \Big( |\tn - \tm|
 \\ \nonumber && \qquad 
     + H(\tn - \tm, t/\tau_R) + H(\tn + \tm, t/\tau_R)
   \Big)\Big]
 \end{eqnarray}
 with
 \begin{equation}
 \label{eq:H}
 H(x,\tau) := \frac{2}{\pi^2} \sum_{p=1}^\infty \frac{1}{p^2}
   \: \cos(p \pi x) \: \big( 1 - \ue^{- \tau p^2} \big).
 \end{equation}

 Our goal is to approximate the function $H(x,\tau)$.  We first note
 that it can be evaluated exactly in the limit $\tau \to \infty$,
 giving
 \begin{eqnarray}
\nonumber
 \label{eq:h0}
 H(x,\infty)
     &=& \frac{2}{\pi^2} \sum_{p=1}^\infty \frac{1}{p^2} \cos(p \pi x)
     = \Big[\frac{1}{2}(x-1)^2 - \frac{1}{6}\Big] \: \mbox{mod} \: 2
  \\ &=:&  \hinf(x) 
 \end{eqnarray}
 To prove this relation, one simply calculates the coefficients of the
 Fourier series of $\hinf(x)$. The leading correction term at $\tau \gg 1$ 
 is the term proportional to $\ue^{-\tau}$, hence we obtain
 \begin{equation}
 \label{eq:H_large}
 H(x,\tau) \approx \hinf(x) - \frac{2}{\pi^2} \ue^{- \tau} \: \cos(\pi x)
 \quad \mbox{at} \quad  \tau \gg 1
 \end{equation}
 
 Likewise, we can evaluate the behavior of $H(x,\tau)$ exactly in the
 limit of $\tau \to 0$ and small $|x|\ll 1$. In that limit, one can
 replace the sum over $p$ by an integral, giving
 \begin{equation}
   H(x,\tau) \approx \frac{2}{\pi^2} \int_0^\infty \!\!\! \ud p \:               \frac{1}{p^2} \:
   \cos(p \pi x) (1 - \ue^{-\tau p^2}) = \sqrt{\tau} \;
   \Phi\big(x/\sqrt{\tau}\big)
 \end{equation}
 \begin{equation}
  \label{eq:phi}
  \mbox{with} \quad
 \Phi(y) = \frac{2}{\pi^2} \; \Big(
   \ue^{- \frac{1}{4} (\pi y)^2} \sqrt{\pi}
   - \frac{\pi^2}{2} |y| \: (1 - \mbox{Erf} (\pi |y|/2))
 \Big),
 \end{equation}
 where Erf is the error function.  Next we seek to approximate
 $H(x,\tau)$ over the whole range $x \in [-2:2]$ for finite small
 $\tau \ll 1$. To this end, we exploit the relation $H(x,\tau) =
 H(|x|,\tau) = H(2-|x|,\tau)$ and make the Ansatz $H(x,\tau) \approx
 H_c(x,\tau) + C(\tau)$ with
 \begin{equation}
 \label{eq:H_c}
 H_c(x,\tau) = \sqrt{\tau}  \: 
   \Big( \Phi\big(\frac{|x|}{\sqrt{\tau}}\big) 
   + \Phi\big(\frac{2-|x|}{\sqrt{\tau}}\big) \Big) 
 \end{equation}
 where we choose the offset $C(\tau)$ such that the approximation
 is best at $x=1$, i.e., $C(\tau) = H(1,\tau)-H_c(1,\tau)$.
 Thus we need to find a good approximation 
 for $H(1,\tau)$ at finite small $\tau$.  We know
 $H(1,0)=0$ and
 \begin{equation}
 \partial_\tau H(1,\tau) 
     = \frac{2}{\pi^2} \sum_{p=1}^\infty (-1)^p \:  \: \ue^{-\tau p^2} 
     = \frac{1}{\pi^2}(\theta_4(\ue^{-\tau})-1) 
     \stackrel{\tau \to 0}{\longrightarrow} - \frac{1}{\pi^2}
 \end{equation}
 where $\theta_4(y)$ is the Theta ''constant''.
 Numerically, it turns out that $\theta_4(\ue^{-\tau})$ closely
 follows
 \begin{equation}
 \theta_4(\ue^{-\tau}) \approx
 2 \sqrt{\frac{\pi}{t}} \ue^{-\pi^2/4t}
 = \pi^2 \partial_\tau H_c(1,\tau)
 \end{equation}
 over a fairly large range of $\tau > 0$. The deviation from the exact
value of $\theta_4(\ue^{-\tau})$ scales roughly as $\ue^{-20/t}$.
 Thus we can approximate
 $\partial_\tau H(1,\tau) \approx - 1/\pi^2 + \partial_\tau H_c(1,\tau)$
 and hence $ H(1,\tau) \approx - \frac{1}{\pi^2} \: \tau +
 H_c(1,\tau)$.  Inserting this into our Ansatz above, we obtain
 \begin{equation}
 \label{eq:H_small}
 H(x,\tau) \approx
 H_c(x,\tau) - \tau/\pi^2
 \quad \mbox{at} \quad  \tau \ll 1
 \end{equation}

\begin{figure}[t]
 \centering
 \includegraphics[width=9cm]{\dir/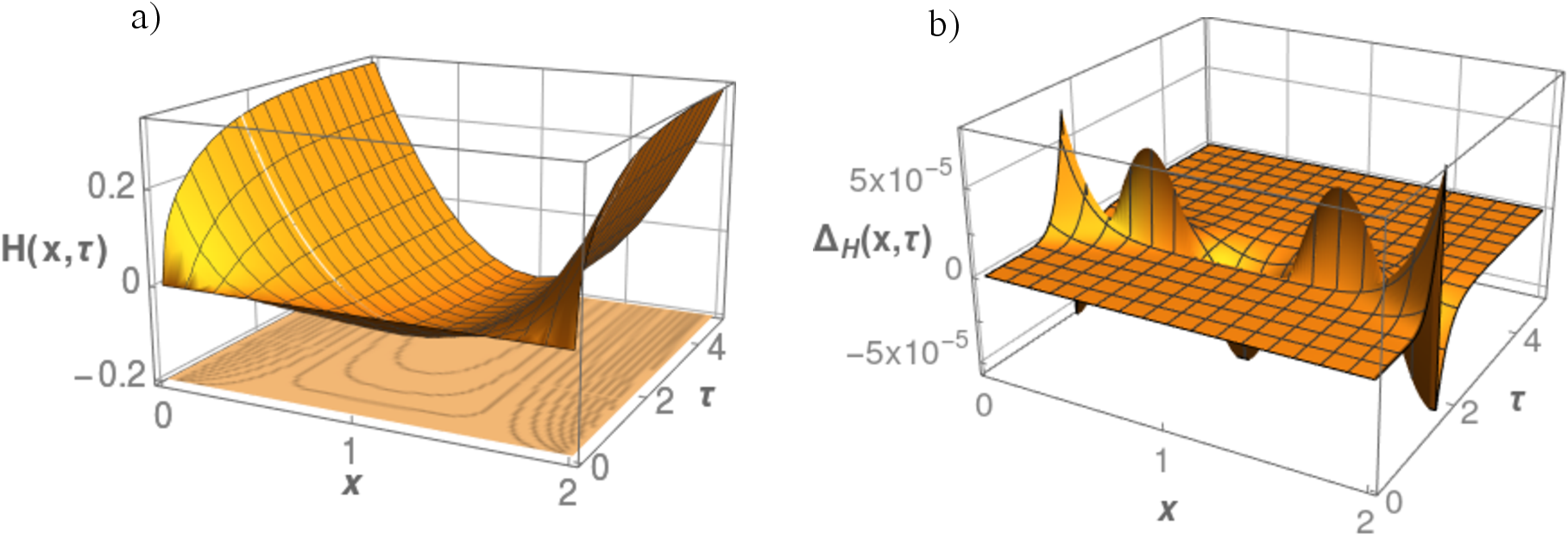}
 \caption{Illustration of (a) $H(x,\tau)$ and (b) deviation
 $\Delta(x,\tau) = \Happr(x,\tau) - H(x,\tau)$
 between the approximate expression, Eq.\ (\ref{eq:Happr}) and the
 numerical value, Eq.\ (\ref{eq:H}), in the range $x \in [0:2]$.
 \label{fig:gqt_appr}
 }
 \end{figure}

 Starting from the two limiting expressions,  Eqs.\
 (\ref{eq:H_small}) and (\ref{eq:H_large}), we now make the
 overall Ansatz
 \begin{equation}
 \label{eq:Happr}
 \Happr(x,\tau) = \left\{
 \begin{array}{ll}
 H_c(x,\tau) - \tau/\pi^2 & : \tau < \tau^* \\
 \hinf(x) - \frac{2}{\pi^2} \ue^{- \tau} \cos(\pi x) & : \tau > \tau^*
 \end{array} \right.
 \end{equation}
 We choose the crossover value $\tau^*=1.7$ such that it minimizes the
maximum absolute value of the deviation $\Delta_H(x,\tau)=
\Happr(x,\tau) - H(x,\tau)$ between the approximate expression
(\ref{eq:Happr}) and the numerical evaluation of the full expression,
Eq.\ (\ref{eq:H}).  With this approach, we can reach
\mbox{$\max(|\Delta_H(x,\tau)|) \le 10^{-4}$} over the whole range of
$x$ and $\tau$.  Figure \ref{fig:gqt_appr} shows $H(x,\tau)$ and the
$\Delta_H(x,\tau)$ as a function of $x$ and $\tau$.

 Inserting $\Happr(x,\tau)$ into Eq.\ (\ref{eq:g}) and performing some
 further simple transformations, we finally obtain Eq.\
 (\ref{eq:g_appr}) in the main text.  From \mbox{$\mbox{max}(\Delta_H)
 \le 10^{-4}$}, we can estimate that the relative error of $g_{\alpha
 \beta}(\qq,t)$ is less than $2 (q \Rg)^2 \times 10^{-4}$ over the
 whole range of $\qq$ and $t$.

\end{appendix}



\bibliography{mobility}




\end{document}